\documentclass[aip,graphicx,amsfonts]{revtex4-1}

\usepackage{graphicx}
\usepackage{color}

\draft

\newcommand{\Eqref}[1]{Equation~(\ref{#1})}
\newcommand{\eqref}[1]{Eq.\,(\ref{#1})}

\newcommand{\figref}[1]{Fig.\,\ref{#1}}
\newcommand{\Figref}[1]{Figure \ref{#1}}

\newcommand{\secref}[1]{\S\ref{#1}}

\newcommand{\appref}[1]{Appendix \ref{#1}}
\newcommand{\p}{\mathcal{P}}

\newcommand{\f}{\overline{\mathcal{F}}}
\newcommand{\n}{\mathcal{N}}

\newcommand{\JL}{J}
\newcommand{\vlambda}{\overline{\lambda}}
\newcommand{\vmu}{\overline{\mu}}

\newcommand{\C}{\overline{C}}

\newcommand{\vsigma}{\overline{\sigma}}

\newcommand{\vI}{\overline{A}}

\newcommand{\BM}{\overline{\overline{M}}}

\newcommand{\etal}{\emph{et. al.}}

\begin{document}

\title{Recent Developments in Bayesian Inference of Tokamak Plasma Equilibria and High-Dimensional Stochastic Quadratures}

\author{G.~T.~von~Nessi}
\email[]{greg.vonnessi@entrolution.net}

\author{M.~J.~Hole}
\affiliation{
Research School of Physical Sciences and Engineering,
The Australian National University,
Canberra ACT 0200, Australia
}

\author{the~MAST~Team}
\affiliation{
EURATOM/CCFE Fusion Association, Culham Science Centre,
Abingdon, Oxon, OX14 3DB, UK
}

\begin{abstract}
We present recent results and technical breakthroughs for the Bayesian inference of tokamak equilibria using force-balance as a prior constraint. Issues surrounding model parameter representation and posterior analysis are discussed and addressed. These points motivate the recent advancements embodied in the Bayesian Equilibrium Analysis and Simulation Tool (BEAST) software being presently utilised to study equilibria on the Mega-Ampere Spherical Tokamak (MAST) experiment in the UK (von Nessi \etal 2012 \emph{J. Phys. A} \textbf{46} 185501). State-of-the-art results of using BEAST to study MAST equilbria are reviewed, with recent code advancements being systematically presented though out the manuscript.

\end{abstract}


\maketitle
\section{Introduction}
The importance of equilibrium magnetic field reconstruction in tokamaks is well understood throughout fusion science.\cite{goedbloed2004, hutchinson2005, bellan2008, goedbloed2010, wesson2011} Indeed, it is the geometry of the equilibrium magnetic field that provides a canonical coordinate, via indexing of nested flux surfaces, which is needed for a wide variety of post shot theoretical and diagnostic data analysis.\cite{hutchinson2005, wesson2011} Equilibrium reconstruction also gives the outer boundary of the plasma: a key element to many open-circuit, real-time control methodologies.\cite{ferron1998, mcardle2004, mazon2010, blum2012} While schemes exist for plasma control using only classical electrostatics to determine the boundary reconstruction\cite{kurihara2000, kurihara2005}, the vast majority of reconstructions of the internal magnetic geometry rely upon solving kinetic force-balance equations with a single solution being chosen as the best fit to available diagnostic data \cite{hutchinson2005, bellan2008, wesson2011}. This approach to internal reconstruction is most famously implemented through the EFIT code (or variants thereof) that uses Picard iteration to find solutions of the Grad-Shafranov (GS) force-balance equation, which best fit data observed from equilibrium magnetic diagnostics (e.g. fluxloops and pickup coils).\cite{lao1985, lao1985b, zwingmann2003} While this approach of leveraging the GS equation to perform equilibrium reconstruction has been successfully utilised throughout the field, the accuracy of the method is intrinsically linked to how accurately the GS equation accounts for all the equilibrium forces in the plasma. Indeed, factors such as flow and isotropy need to be explicitly added into the underpinning force-balance equations to be correctly accounted for in the equilibrium reconstruction.\cite{fitzgerald2013} Moreover, solutions to equilibrium reconstruction are not generally unique\cite{lao1985, lao1985b}; and thus, experiment-specific numerical schemes are frequently employed to guarantee that the Picard iteration converges to a physical solution.

In parallel to the inclusion of more physics in equilibrium solvers, there has been the improvement in the diversity, accuracy and resolution of plasma diagnostics. Interpretation, however, often requires a detailed knowledge of the plasma equilibrium. For example, inference of the toroidal current profile $j_\phi(\psi)$ from line of sight measurements of the polarisation angle requires a knowledge of the poloidal flux $\psi$ across the plasma. Formally, diagnostic forward functions relate the vector of plasma parameters $\vlambda$ to the measurement vector $\vmu$. For a linear system, such as toroidal current inference in a double null configuration,  $\vlambda$ and $\vmu$ are normally related through a response matrix $\BM$ with additional contributions $\C$, such that $\vmu = \BM \vlambda + \C$. Inference, or parameter estimation, involves inverting this relationship to give plasma parameters $\vlambda$ that are consistent with the data $\vmu$. A widespread technique used is least-square fitting, used for instance in EFIT, in which prior assumptions are included via a penalty term in the fit. 

Given the large data-sets and complicated models, an arguably more rigorous approach to the integrated data-modelling challenge is the Bayesian approach to inference in fusion plasmas.
In contrast to least square fitting, the Bayesian approach to inference in fusion plasmas, developed by multiple authors,  \cite{cottrell1991, millar2000, fischer2003, dinklage2004, svensson2004, svensson2008, ford2008, reginatto2008} involves the specification of an initial prior probability distribution function (pdf), $P(\vlambda)$, which is then updated by taking into account information that the measurements provide through the likelihood pdf $P(\vmu|\vlambda)$. The result is the posterior distribution $P(\vlambda|\vmu)$ given by Bayes' formula 

\begin{equation}
P(\vlambda|\vmu) = \frac{P(\vmu|\vlambda)P(\vlambda)}{P(\vmu)}.\label{eq:Bayes}
\end{equation}

The advantage of the Bayesian approach over traditional inversion techniques is two-fold: (i) prior knowledge, including known parameter inter-dependencies is made explicit, and (ii) as the formulation is probabilistic, random errors, systematic uncertainties and instrumental bias are an integral part of the analysis rather than an afterthought. 

The application of Bayesian approach to inference and parameter estimation in complex physics problems is not new, with fields ranging from astronomy to nuclear reaction analysis \cite{dose2003,vontoussaint2011}.  A topical illustrative example comes from parameter estimation in the climate science community, in modelling land-surface-atmosphere processes and global carbon dioxide concentrations in the atmosphere.\cite{enting1995, wang2009} The models for carbon dioxide exchange are complex and span more equations of state than a plasma. The Community Atmosphere Biosphere Land Exchange model (CABLE) is a land surface model, \cite{wang2011, haverd2013} used to calculate the fluxes of momentum, energy, water and carbon between the land surface and the atmosphere and to model the major biogeochemical cycles of the land ecosystem. It solves radiation, heat and mass flow transport on a global scale, accounting for many different land ecosystems. Data is disparate and vast, and comes from an flux towers, carbon stock, carbon in biomass, litter falls, meteorological data, stream flow and satellite imagery.\cite{haverd2013} In this community, the challenge of model and data integration, also called model--data fusion or model--data synthesis, is defined
as combining models and observations by varying some properties of the model, to give the optimal combination of both.\cite{raupach2005} The topic of model--data fusion is crucial to give credibility to the calculation of carbon dioxide fluxes and processes in the atmosphere, and thus provide a reliable basis for public policy on climate change. Bayesian inference, together with other model-data fusion techniques, is extensively utilised. 

In contrast to climate science, the systematic inclusion of uncertainties in both data and models has, to-date, not been a strength of the fusion community. Several facets are driving change. ITER discharges will be extremely expensive, and so it will be crucial to maximise the value of acquired data. The challenging environment of a fusion reactor will mean fusion power plants will operate with a very much reduced set of diagnostics. Finally, as more physics is added to force-balance descriptions, there is a need to validate physics models. Once validated, such 
models may be able to be used as a constraint in equilibrium reconstruction to infer additional information about the plasma, and thereby create ``model diagnostics''. These aspects have motivated the recent development of a Bayesian approach to equilibrium reconstruction\cite{svensson2008, hole2010, hole2010a, hole2011a, vonnessi2012}, with one line of research producing a code called the Bayesian Equilibrium Analysis and Simulation Tool (BEAST), which is able to quantify fit degeneracies and infer spatially-localised discrepancies from a force-balance solution\cite{vonnessi2013}. This paper presents further research advancements since the introduction of BEAST by the authors\cite{vonnessi2013} and that have subsequently been used to advance the code.

The paper is structured as follows: \secref{Sec1} gives a brief overview of Bayesian inference and its application to equilibrium reconstruction.  This is followed by a general discussion on the computational challenges surrounding Bayesian equilibrium reconstruction and how these have been addressed by recent advancements, coded into BEAST. State-of-the-art results coming from the use of BEAST to analyse discharges on the Mega-Ampere Spherical Tokamak (MAST) are then presented, followed up be a concluding remarks encompassing future research endeavours and a summary of the current status of BEAST. Finally, two appendices detail specifics on recent advancements surrounding posterior optimisation and integration.

\section{Brief Review of Bayesian Inference in Fusion Plasmas}\label{Sec1}

Bayesian inference offers an alternate approach to equilibrium modelling in fusion plasmas \cite{cottrell1991, fischer2003, svensson2004, ford2010, hole2010}, and a pathway to validate different equilibrium model descriptions \cite{hole2011a, vonnessi2013}.  Some understanding can be gleaned by understanding the application of Bayes' theorem to a single observation with $\mu_i$ and $\sigma_i$. In this case, Bayes' formula becomes
\begin{equation}
\p(\vlambda|\mu_i,\sigma_i)\propto\p(\mu_i|\vlambda,\sigma_i)\p(\vlambda),\label{eq2}
\end{equation}
where $\vI$ has been dropped to simplify the notation; this convention will be maintained throughout the remainder of the paper. As $\vmu$ and $\vsigma$ are given and thus assumed to be constant, so is $\p(\vmu,\vsigma)$, which is reflected by the proportionality in \eqref{eq2}. The \emph{forward model}, $\f(\vlambda)$, is implicitly contained within $\p(\mu_i|\vlambda,\sigma_i)$ and is a deterministic mapping from the space of model parameters to the space of associated diagnostic observations. That is, the forward model generates a prediction of what the diagnostic observations would be, given a set of model parameters.

In most treatments likelihoods are assumed to be of the form
\begin{equation}
\p(\mu_i|\vlambda,\sigma_i)=\n(\mu_i-\mathcal{F}_i(\vlambda),\sigma_i^2),\label{eq2a}
\end{equation}
where $\n$ is represents a Gaussian distribution over pair-wise independent variables. The first argument of the Gaussian distribution represents the mean vector, with the second being the entries in a diagonal covariance matrix. The justification for the form of the likelihood is discussed elsewhere\cite{jaynes2003,sivia2006,svensson2008,vonnessi2013}. 

Using the likelihood in \eqref{eq2a}, the following form can be written for the posterior:
\begin{equation}
\p(\vlambda|\vmu,\vsigma)\propto\left(\prod_i\n(\mu_i-\mathcal{F}_i(\vlambda),\sigma_i^2)\right)\p(\vlambda).\label{eq3}
\end{equation}
From \eqref{eq3}, it is clear that the posterior represents a probability distribution over model parameters, if given a set of diagnostic observations and uncertainties. \Eqref{eq3} is the form which is ultimately integrated to find statistical moments of model parameters and various marginalisations thereof.

Hole \etal\ have implemented Bayesian inversion on MAST using the MINERVA framework.\cite{svensson2007} Within this framework, probabilistic graphical models are used to project the dependence of the posterior distribution function on the prior, the data, and the likelihood.  An advantage of this approach is that it visualises the complex interdependency between data and model, and thus expedites model development.  The techniques of Bayesian inference have also been inverted to provide
a tool to check data consistency.\cite{vonnessi2012}

Various authors have developed Bayesian inference techniques for fusion plasmas that combine information from a wealth of diagnostics to enable probabilistic calculation of plasma configuration, \cite{hole2010} provide automatic identification of faulty diagnostics,\cite{vonnessi2012} and developed a  validation tool for generalised force-balance models \cite{hole2011a}. Critically, Bayesian techniques propagate experimental uncertainty correctly, and enable the relative uncertainty between acceptable physical models to be quantified. In von Nessi \etal \cite{vonnessi2013}, a new method, based on Bayesian analysis, is presented which unifies the inference of plasma equilibria parameters in a tokamak with the ability to
quantify differences between inferred equilibria and GS solutions. At the heart of this technique is the new concept of weak observation, which allows multiple forward models to be associated with a single diagnostic observation. This new idea subsequently provides a means by which the space of GS solutions can be efficiently characterised via
a prior distribution. The posterior evidence (a normalisation constant of the inferred posterior distribution) is also inferred in the analysis and is used as a proxy for determining how relatively close inferred equilibria are to force-balance for different discharges/times.

\begin{figure}[!htb]
\includegraphics[width=\textwidth]{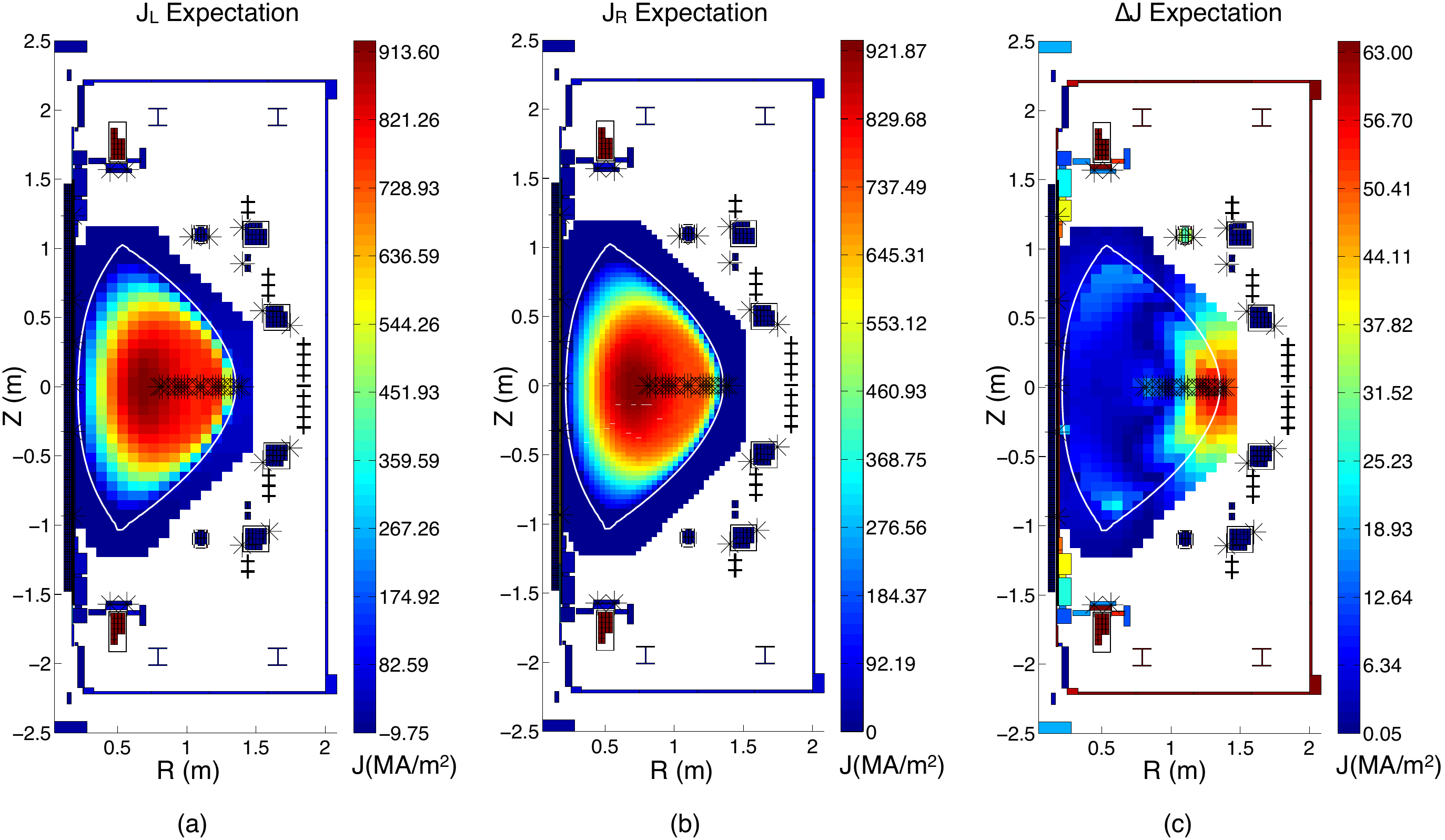} 
\caption{\label{fig:22254}
Expectation values of $J(R, Z), J_{GS} (R, Z)$ and $ \Delta J(R, Z)$ inferred for MAST discharge \#22254 at 350 ms, as calculated from 1800 samples of the posterior, using pickup coils, flux loops, MSE and Rogowski coil data. The inferred last closed flux surface is indicated in white on each figure. Flux loop locations are indicated by stars outside the plasma region; position and orientation of pickup coils are indicated via heavy bars on the out-board edge of the first wall and as a vertically oriented column line along the solenoid; and MSE observation positions are indicated by the stars across the mid-plane inside the plasma region. Panel (a) shows $J(R, Z)$ current density data, with the current densities in (b) reflecting that of $J_{GS} (R, Z)$. Note that the number and size of beams representing $J(R, Z)$ and $J_{GS} (R, Z)$ are allowed to differ in BEAST inferences. (c) shows the magnitude of the current density difference as averaged across each 2D rectangular step corresponding to $J(R, Z)$. Reproduced with permission from Fig. 2 of von Nessi and Hole\cite{vonnessi2013}.}
\end{figure}

Figure \ref{fig:22254} shows expectation values of the toroidal current density inferred from (a) a toroidal current beam model, (b) a GS constraint, in which $J_\phi$ is computed from GS from a $\psi$ surface, together with fits to the pressure and toroidal flux function, and forward models for magnetics, total plasma current and MSE predictions, and (c) the difference between the two. The difference in $J_\phi$ can give some indication to physical effects neglected in the GS equation, and/or reflect diagnostic disagreement. In this case the discrepancy is largest at the outboard mid-plane, and of order of 10\%. Using nested sampling, it is possible to integrate over the evidence, and thus compute $\ln(P(\vmu))$ of
the inferred hyper-parameter, $\sigma^2$, which is the average current variance between GS and toroidal current beam values. The smaller the value of $\ln(P(\vmu))$, the larger the degree of freedom necessary to predict diagnostic observations relative to other cases. For 22254 at 350~ms $\sigma_*^2 =9.46 \times 10^{-3} \pm 1\times 10^{-5}$~(kA)$^2$ and $\ln(P(\vmu))=765$, while $\sigma_*^2 = 0.2034 \pm 2\times 10^{-4}$~(kA)$^2$ and $\ln(P(\vmu)) = -39.0$  for adjacent discharge \#24600 at 265ms. This meant \#22254 was much closer to GS, and/or had fewer diagnostics in conflict, than \#24600.

\section{Mathematical and Computational Challenges}\label{Sec2}
The equilibria inference described in \secref{Sec1} poses a number of unique computational challenges when it comes to analysing the associated, high-dimensional (i.e. having more than 1000 dimensions) posterior distribution. This section discusses emergent points and recent research pursuits surrounding the computational aspects of Bayesian equilibrium reconstruction, some of which have led to recent advances in the BEAST code beyond its original introduction in von Nessi \etal\cite{vonnessi2013} 

\subsection{Plasma Current Representation}\label{SMO}
The beam model used to represent the toroidal plasma current in the BEAST code, typically uses 524 model parameters to simulate a MAST discharge.\cite{vonnessi2012, vonnessi2013} This high dimensionality alone constitutes a significant computational challenge in analysing the associated posterior distribution, as no efficient, general means exist to sample from such distributions.\cite{mackay2003} However, the plasma beam model obviously imposes no intrinsic spatial correlation between cross-sectional points contained within different beams, i.e. without the presence of an informative prior. 

Alternative, more compact (i.e. potentially more computationally efficient), representations for the beam currents have been trialled. Specifically, both a 2D Fourier and 2D Bessel-Fourier representations have been investigated. Neither produced a computational procedure that could achieve the levels of accuracy of the beam model. This outcome is not surprising, given the non-linear nature of the force-balance constraint and the fact that there are no strict symmetries in the plasma current. Indeed, these two points eliminate many paths by which a more compact representation of the beam currents could be achieved, under a force-balance constraint.

%Thus, associated aliasing affects need not be considered, beyond the resolution of the beam grid itself, when interpreting BEAST results. This point embodies the reason why a lower dimensional representation is not simply projected into the high-dimensional plasma beam representation to simplify the computational analysis of the posterior; e.g. representing the current via a low-order, 2D Fourier decomposition and then using a higher order quadrature to construct the appropriate Green's functions for the geometry. Moreover, attempts to lift lower dimensional representations of the toroidal plasma current into the beam model have all consistently lead to poorer fits to diagnostic data and greater deviations from force-balance.

\subsection{Local Maxima and ``Screening'' Solutions}\label{SMO2}

Between flux loops, pickup coils and MSE, there are about 150 diagnostic observations available for equilibrium reconstruction for a typical MAST discharge. Thus, fitting the parameters of the beam model above constitute an underdetermined problem, neglecting any priors. This underdetermined nature affords many ``screening" solutions to exist, where only currents nearest to diagnostic observation points need to be adjusted to compensate for any, otherwise arbitrary, configuration of beam currents. This translates into the posterior having many local maxima, with the global maxima--called the maximum of the posterior (MAP)--generally not corresponding to a physically realistic plasma configuration. While the addition of a force-balance prior serves to greatly reduce the number of these local minima (in addition to making the global maxima correspond to a physically realistic plasma configuration), finding this global maxima is still computationally difficult and constitutes the majority of the computational time in BEAST inferences.\cite{vonnessi2013} Indeed, even with the inclusion of a force-balance constraint, many screening solutions (i.e. local maxima) are still present, with many lying in close proximity to the global maxima. A significant increase in accuracy in inferring the MAP has been achieved through the development of a new non-linear optimisation algorithm, outlined in \appref{SMO3}. This optimiser is based on the Hookes and Jeeves algorithm\cite{hooke1961} but has been heavily modified to avoid screening solutions, when exploring the posterior. Thus, we call this algorithm the ``Screening Mitigation Optimiser'' (SMO).

\subsection{Posterior Marginalisation and Analysis}
The posterior distribution associated with BEAST equilibrium reconstruction is high-dimensional and non-Gaussian, having the majority of the probability ``mass'' in a highly-localised region of model parameter space.\cite{vonnessi2013} Sampling from such distributions is inherently problematic and extremely computationally intensive.\cite{mackay2003} Indeed, Markov-Chain Monte-Carlo (MCMC) methods are too inefficient to employ, as there is little chance for the chain to find (and subsequently stay in) the region of high probability density. Moreover, it is difficult to find bounds on the accuracy of an analytic approximation of the posterior. Thus, BEAST uses a statistical quadrature to build up moments of the posterior directly, rather than approximating these moments through sampling statistics.

The method currently employed by BEAST to integrate the posterior is a generalisation of the modified Nested Sampling\cite{sivia2006,skilling2006} (NS) algorithm presented in von Nessi \etal\cite{vonnessi2013}, called the Stochastic Lebesgue Quadrature (SLQ). SLQ was developed to alleviate inefficiencies in employing MCMC techniques to generate prior samples under a likelihood constraint, which is intrinsically required by NS\cite{sivia2006,skilling2006}. Moreover, SLQ is not affected by possibly ambiguities in classifying a PDF in the inference as part of the likelihood or a prior. SLQ is general enough to work with inferences that even use uninformative priors.

Generally, the method works be approximating the set $\{\vlambda\,|\,\p(\vlambda|\vmu,\vsigma)>C\}$ for any given $C>0$ by a collection of pairwise disjoint hypercubes. These hypercubes are generated from an evolving swarm of model parameter vectors, each of which is already guaranteed to satisfy the given posterior constraint. Extracting a uniform sample from the union of hypercubes is a fast computation that is not only leveraged to evolve the swarm at each step but also provides the statistical basis for the construction of any posterior quadrature. The details of this method are explained in \appref{SLQ}.

Finally, SLQ has recently been deployed in BEAST and has resulted in more thorough exploration of the posterior during quadrature construction, which has ultimately lead to more consistent results coming from the computations (see \secref{Results} for more details). This has been achieved while maintaing the same to slightly shorter computational times, relative to those reported in von Nessi \etal\cite{vonnessi2013}.

\section{Results for Two MAST Discharges}\label{Results}

\begin{figure}[!htb]
\includegraphics[width=\textwidth]{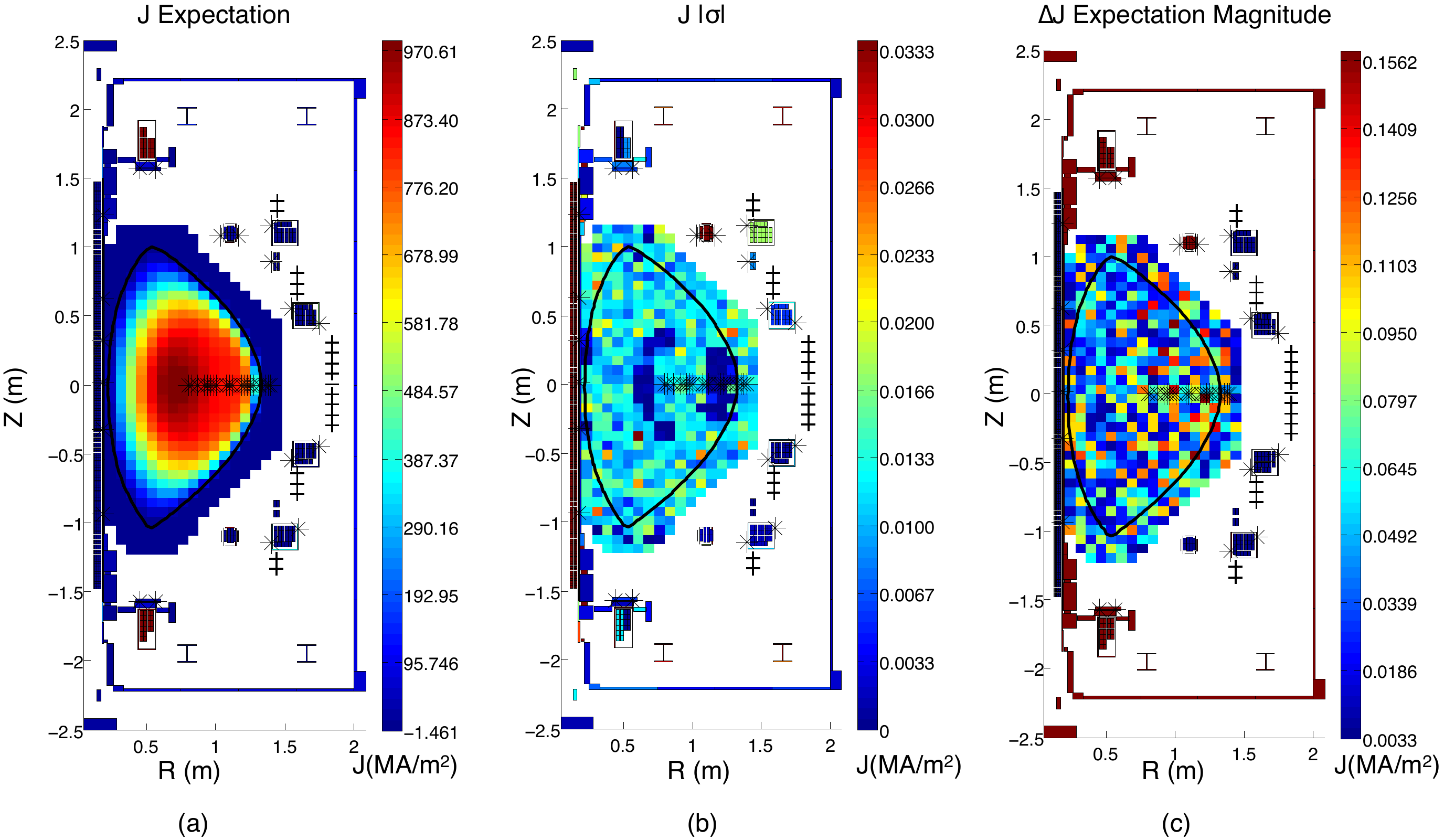}  
\caption{\label{fig::22254BeamData}Expectation and standard deviation values of $\JL(R,Z)$, along with the expectation values of $\Delta J(R,Z)$ inferred for MAST discharge \#22254 at 350ms, as calculated from BEAST, using pickup coils, flux loops, MSE and Rogowski coil data. In each subfigure the LCFS, as inferred from BEAST, is drawn in black. Flux loop locations are indicated by stars outside the plasma region; position and orientation of pickup coils are indicated via heavy bars on the out-board edge of the first wall and as a vertically oriented column line along the solenoid; and MSE observation positions are marked by the stars across the mid-plane inside the plasma region. (a) shows the expectation of $\JL(R,Z)$: the current density data, with (b) presenting the magnitude of one standard deviation thereof. (c) shows the magnitude of the $\Delta J(R,Z)$ expectation (introduced in von Nessi \etal\cite{vonnessi2013}), which directly correspond to local deviations from force-balance, as dictated by the GS equation, with larger magnitudes reflecting a larger deviation.}
\end{figure}

Here we present results from two MAST discharges, which demonstrate BEAST's growth in capabilities since being initially introduced. The discharges analysed were \#22254 at 350ms and \#24600 at 280ms. Both are DnD plasmas, with the former being in H-mode and the latter in L-mode. Discharge \#22254 was part of a hybrid scenario study carried out in MAST and is heated with 3.13MW of NBI power. Contrasting this is \#24600 that was part of an L-mode study being injected with 3.35MW of NBI power. Discharge \#22254 was studied in von Nessi \etal\cite{vonnessi2013} and is revisited here to show how the inference has been improved with recent advancements in BEAST. We look at \#24600 at 280ms, a time shortly after one of the two NBI beams disrupts, to study the impact of NBI disruption on the equilibrium. The following results are obtained from 76 pickup coils, 24 flux loops and 31 MSE observations. Finally, additive bias corrections and conducting surface currents are inferred in every BEAST inference\cite{vonnessi2012,vonnessi2013}; however, these are treated as nuisance parameters, as they do not typically impact the physics interpretation of the results and thus, will not be reported here.

To interpret the results below, we note that BEAST outputs a cross-sectional quantity, $\Delta J(R,Z)$, which indicates how close an associated configuration is to axisymmetric force-balance, with smaller values indicating configurations being relatively closer to force-balance.\cite{vonnessi2013}. Qualitatively, $\Delta J(R,Z)$ reflects the level of discrepancy between the toroidal current density, calculated from the GS equation (which ultimately uses pressure, poloidal current and toroidal current model parameters) and that calculated directly from the plasma beam model; for more details see von Nessi \etal\cite{vonnessi2013}. Thus, relatively large values of $\Delta J(R,Z)$ can be viewed as an indicator for missing physics in the force-balance model.

\subsection{\#22254 at 350ms}

\begin{figure}[!hbt]
\includegraphics[width=.666\textwidth]{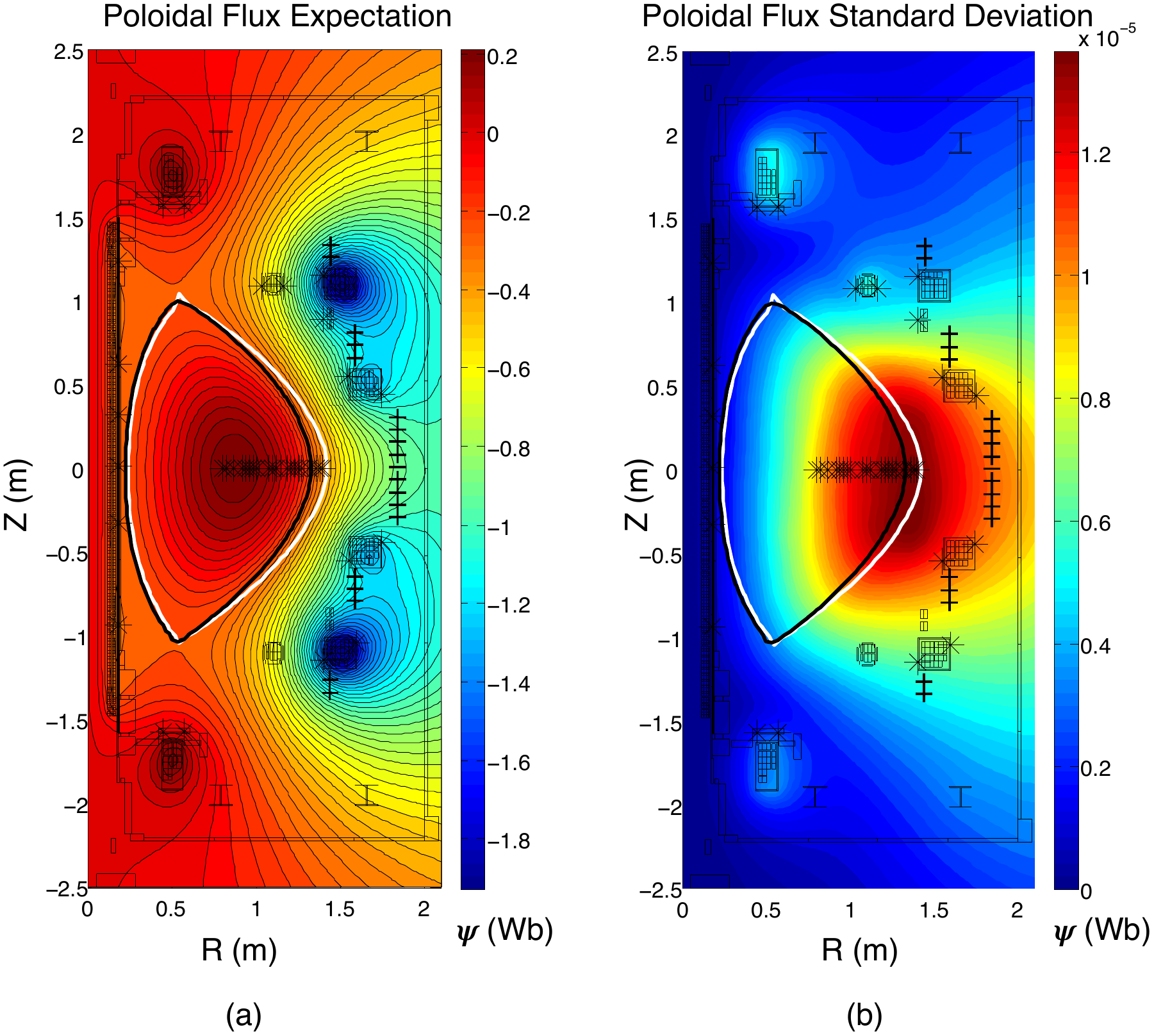}  
\caption{\label{fig::22254PoloidalFlux}Poloidal flux function expectation and standard deviation as calculated by BEAST for \#22254 at 350ms. Positions of magnetics and MSE observation points are indicated as they were in \figref{fig::22254BeamData}. In both subfigures the EFIT LCFS is plotted in white with the BEAST LCFS overlaid in black.}
\end{figure}

In \secref{Sec2} the SMO algorithm was introduced, which has consistently found diagnostic fits that were closer to force-balance than results coming from other optimisers. This is exemplified in \figref{fig::22254BeamData}, where a fit for \#22254 is found with a $\Delta J(R,Z)$ with values $\sim 500$ times smaller than the initial results presented in von Nessi \etal\cite{vonnessi2013}, which are reproduced in \figref{fig:22254}. In particular, a force-balance solution was able to be much better reconciled on the outboard edge of the plasma around the MSE measurements. Ultimately this has resulted in a retraction of the plasma boundary compared to the EFIT LCFS (shown in \figref{fig::22254PoloidalFlux} for comparison), which is only constrained by flux loops and inboard pickup coils, not MSE. The difference in plasma volume accounts largely for the discrepancy of $\beta_p+\frac{\ell_i}{2}$ between EFIT and BEAST: $1.08$ and $0.9821\pm .0008$, respectively, with $95\%$ confidence intervals on the BEAST result. This small uncertainty in the BEAST result coincides with the inference being over-determined (i.e. a very small degree of degeneracy), when a force-balance prior is leveraged against the unbiased space of model parameters. This makes sense, as the GS equation is an elliptic, semi-linear PDE having unique solutions\cite{gilbarg2001}, which is paramaterised only by pre-defined representations of the pressure and poloidal current profiles (polynomials of degree 3 and 5 respectively for these results). Thus, the space of all configurations is biased toward an eight-dimensional submanifold, on which the problem becomes over determined, when reconciled against over 100 diagnostic observations. One may argue that the boundary paramaterization also needs to be accounted for; but this can be determined independently of solving the GS equation\cite{kurihara2000, kurihara2005} and does not embody genuine degrees of freedom in the inference.

\Figref{fig::22254BeamData}(b) shows the magnitude of a single standard deviation for the current density distribution, which are noted to be uniformly much smaller compared to the expectation values in \figref{fig::22254BeamData}(a). This is also consistent with a small uncertainty in the $\beta_p+\frac{\ell_i}{2}$ for the BEAST result.

\Figref{fig::22254PoloidalFlux} shows the poloidal current expectation and first standard deviation magnitude. Here again, the uncertainties are much smaller, as compared to those presented in von Nessi \etal\cite{vonnessi2013}. In addition to, the uncertainty being on the order of five times smaller, the area of greatest uncertainty is larger, being spread across the outboard edge of the plasma, as opposed to be consolidated around the PF coils in von Nessi \etal\cite{vonnessi2013}. The very small uncertainty in the poloidal flux reflects a very high precision in flux-surface positions for a GS model of force-balance. The expectation value of the poloidal flux function is very similar to the results in von Nessi \etal\cite{vonnessi2013}, with the biggest difference being that the outboard LCFS has slightly migrated toward the core of the plasma.

For \#22254 at 350ms, the inferred pressure, poloidal current and q-profiles were all inferred with very similar expectations and uncertainties, compared to previous results\cite{vonnessi2013}. In general, these profiles exhibit expectations that are in good agreement with EFIT and have extremely small uncertainties. Moreover, these profile appear to be close to Gaussian marginalisations, showing symmetric uncertainties and having their expectations coincide with their respective MAPs.

\subsection{\#24600 at 280ms}

\begin{figure}[!htb]
\includegraphics[width=\textwidth]{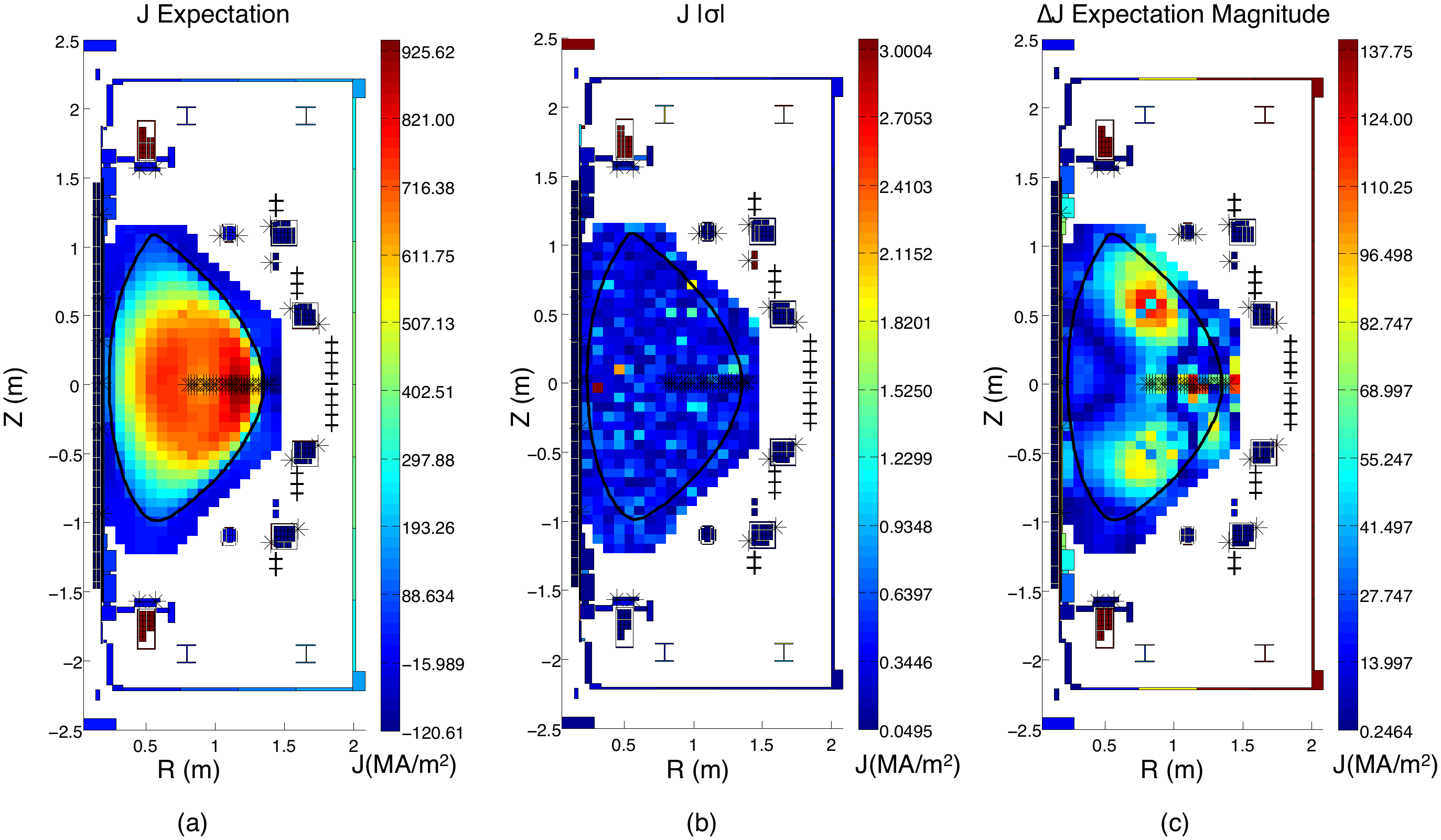}  
\caption{\label{fig::24600BeamData}Expectation and standard deviation values of $\JL(R,Z)$, along with the expectation values of $\Delta J(R,Z)$ inferred for MAST discharge \#24600 at 280ms, using the same diagnostics as in \figref{fig::22254PoloidalFlux}. The LCFS and diagnostic positions are also likewise indicated. (a) shows the expectation of $\JL(R,Z)$, with (b) again showinging the magnitude of one standard deviation thereof. (c) shows the magnitude of the $\Delta J(R,Z)$ expectation.}
\end{figure}

Reflecting 280ms immediately following a NBI disruption, \figref{fig::24600BeamData} shows an equilibrium inference that is significantly out of force-balance. The force-balance discrepancy peaks out around $14.8\%$ at four, spatially separated point, clearly indicated in \figref{fig::24600BeamData}(a). Moreover, the uncertainties on the toroidal current are generally one to two orders magnitude greater than those for \#22254 at 350ms. This relative increase in uncertainty is due to an increase in fit degeneracy, as more degrees of freedom will emerge the farther away from force-balance the inference gets. Interpreting $\beta_p+\frac{\ell_i}{2}$ between BEAST and EFIT is challenging in this context, as the BEAST inference is not in force-balance. However, comparing the EFIT and BEAST values of $1.034$ and $0.894\pm.009$, respectively, for this quantity show that both agree that this value be less about the same amount, when compared to the results from \#22254. Moreover, the uncertainties on the BEAST result are about an order of magnitude greater, which is consistent with the arguments put forth above.

\begin{figure}[!hbt]
\includegraphics[width=.92\textwidth]{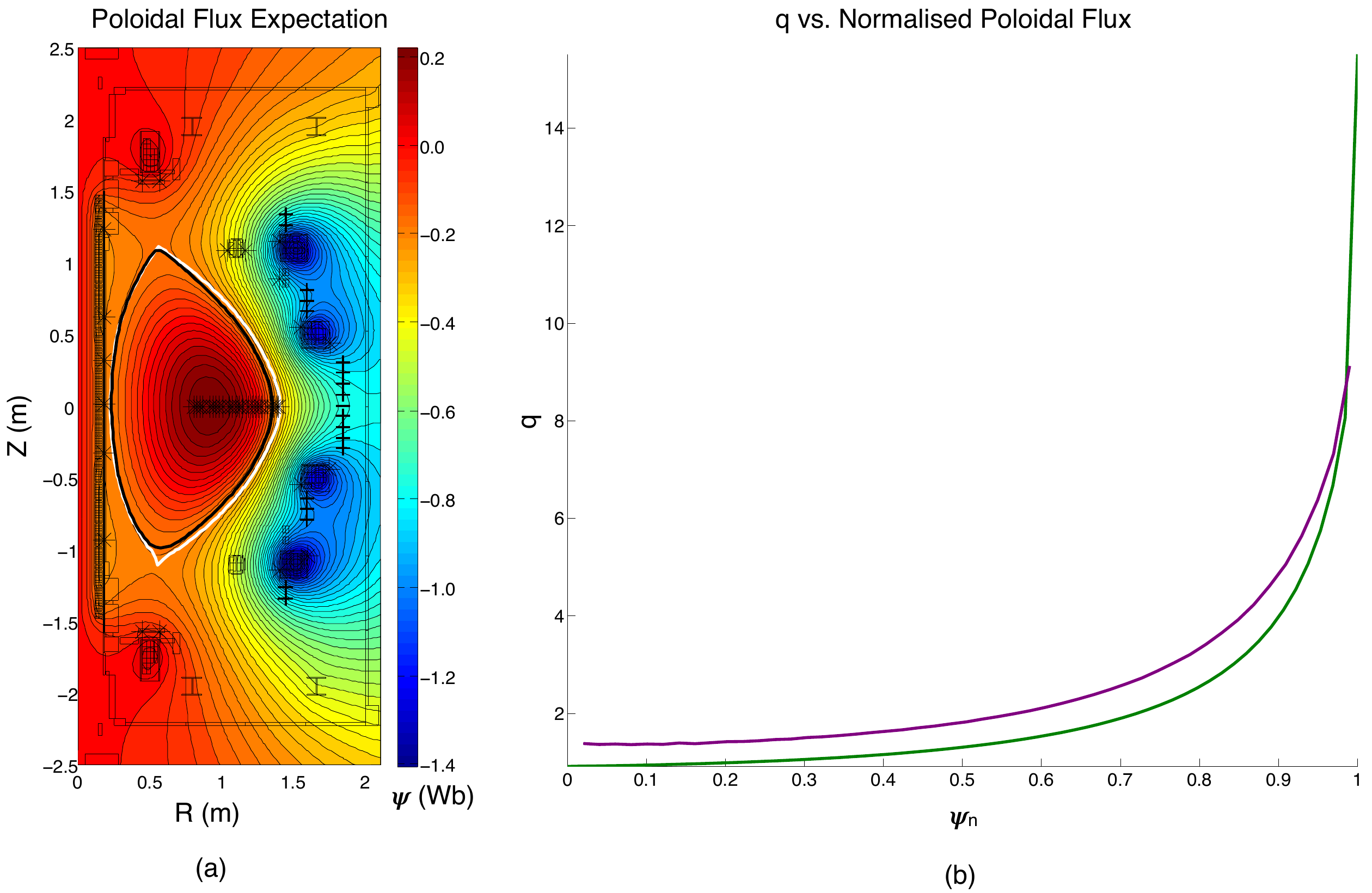}  
\caption{\label{fig::24600PoloidalFlux}Poloidal flux function and q-profile expectations calculated by BEAST for \#24600 at 280ms. Positions of magnetics and MSE observation points in (a) are indicated as they were in \figref{fig::22254BeamData}. In (a), the EFIT LCFS is plotted in white with the inferred LCFS overlaid in black. (b) displays the q-profile as calculated by BEAST (the expectation) and EFIT, represented by the purple and green lines, respectively. Uncertainties on the BEAST q-profile are too small to visually resolve on the scale of the figure and have thus been suppressed.}
\end{figure}

\Figref{fig::24600PoloidalFlux}(a) shows a good agreement between both the EFIT an BEAST LCFS. Again, this can be explained by the arguments in the preceding paragraph regarding the growth of degeneracy uncertainty and recalling that the boundary can be inferred independently of force-balance constraints. However, \figref{fig::24600PoloidalFlux}(b) shows a discrepancy of about $0.5$ between the q-profiles of EFIT and BEAST. This is mostly due to the fact that BEAST's q-profile is strongly constrained to MSE measurements, while the EFIT reconstruction is not. This discrepancy in q-profile coincides with the difference between the BEAST and EFIT inferences of $\beta_p+\frac{\ell_i}{2}$ (despite both plasmas having similar volumes), as $\ell_i$ depends on magnetic field geometry.

\begin{figure}[h]
\includegraphics[width=.75\textwidth]{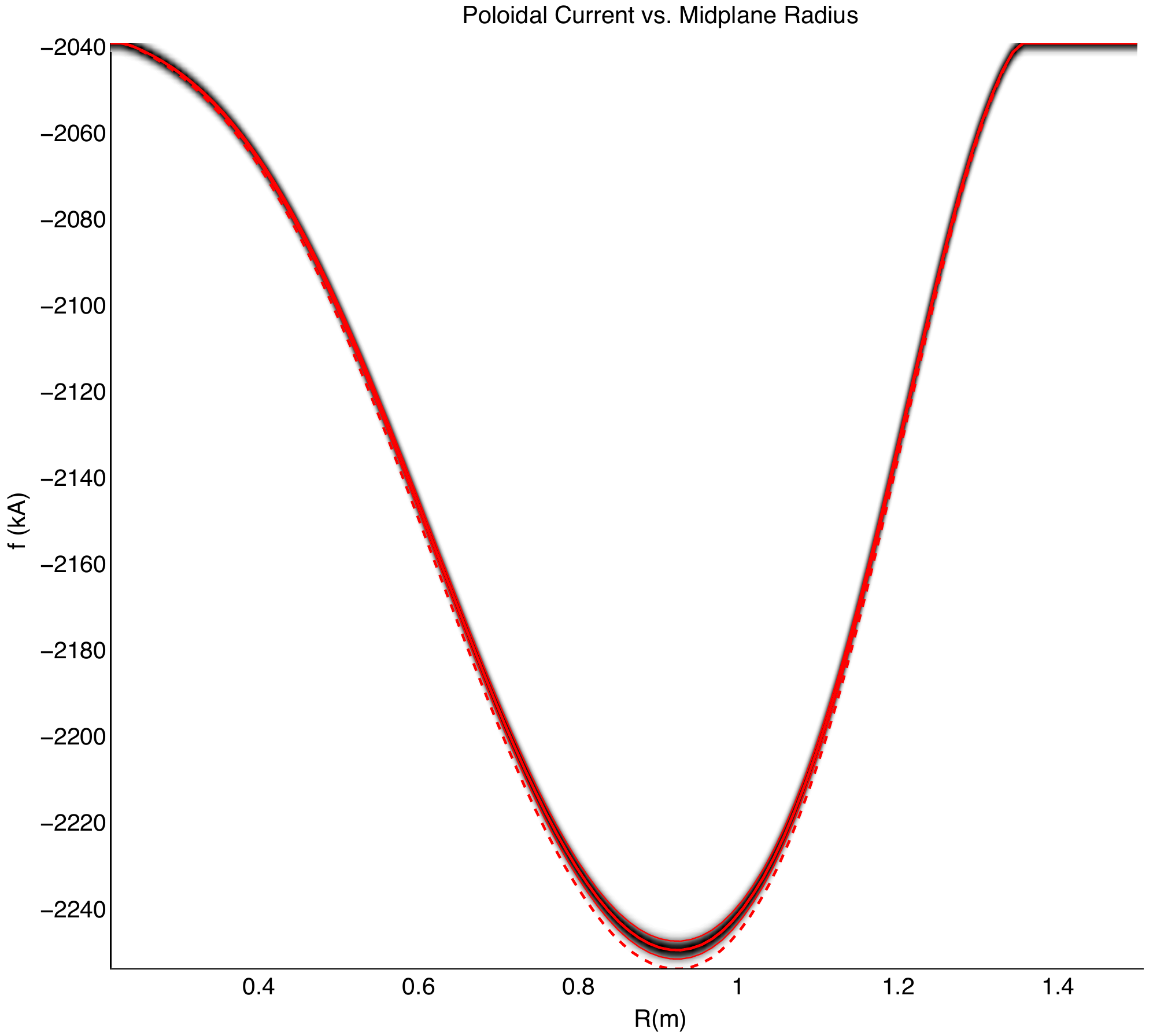}  
\caption{\label{fig::24600pandf}Inferred poloidal current for \#24600 at 280ms, right after the southwest neutral beam disrupts. The dotted line indicates the MAP profile, with the thick line being the expectation. The thin lines represent upper/lower confidence intervals of 95\%, with the shading posterior probabilities of quadrature points.}
\end{figure}

In \figref{fig::24600pandf}, the profile for the poloidal current is shown, demonstrating the non-Gaussian nature of the quantity. Indeed, the plot shows the MAP of the profile lying outside the $95\%$ confidence intervals, surrounding the expectation, implicating the profile as highly non-Gaussian in the core region of the plasma. This result demonstrates BEAST's ability to resolve non-Gaussian structures in even high-dimensional marginalisation of the posterior. Echoing the discussion put forth in von Nessi \etal\cite{vonnessi2013}, we ascribe no rigorous physical interpretation to the kinetic pressure, as the inference is far from force-balance and there exists no direct constraint on the kinetic pressure in the inference. Thus, we do not present the pressure profile for this inference here.

\subsection{Information Theoretic Scalars}

BEAST routinely outputs various information theoretic scalars, such as the evidence and relative entropy between posterior and prior distributions. However, interpreting the meaning of these quantities, outside the realm of model comparison, becomes difficult for the following reasons. First, it is well easily understood that likelihoods are not probability distributions\cite{mackay2003}; and even in the form of \eqref{eq2a}, likelihoods still enjoy a gauge freedom corresponding to an arbitrary scalar multiplier, which will directly affect the value of the evidence. Moreover, the number of observations itself will also have an obvious impact on the evidence (c.f. \eqref{eq3}). 

When leveraging implicit techniques to construct priors, like the methods employed in BEAST to bias toward force-balance, faithfully calculating the relative entropy between prior and posterior distributions is difficult, as the prior is not normalised during the quadrature construction. Indeed, we only need to leverage relative probabilities from the prior to construct the posterior quadratures, when using a technique like SLQ. It is possible to classify the force-balance prior as part of the likelihood in this situation, but this leads to ambiguities as to how to classify distributions as priors or likelihoods. Given this, we instead report the relative entropy, $\mathcal{S}_p$, between posterior and the approximating uniform distribution used in the SLQ calculation (see \appref{SLQ}). To give some context to the meaning of $\mathcal{S}_p$, the volume of the approximating uniform distribution, $|\mathcal{B}|$, is reported via its natural logarithm, $\ln|\mathcal{B}|$, to be consistent with the notation in \appref{SLQ}. 

For \#22254 at 350ms the relative entropy between the posterior and the initial approximating uniform distribution having $\ln|\mathcal{B}| = 2403.6$ was $\mathcal{S}_p = 29.7\pm0.8$ bits, with the uncertainty being the $95\%$ confidence interval. Discharge \#24600 at 280ms had $\mathcal{S}_p = 42.0\pm1.3$ bits relative to an initial uniform distribution having $\ln|\mathcal{B}| = 2327.3$. Generally speaking, these values reflect how much information was provided by both diagnostic observations and the force-balance prior in the inference. As uncertainties were generally higher for \#24600 at 280ms, a higher relative entropy means that the observations and prior were more effective at excluding outlier configurations, relative to those for \#22254. Thus, while $\#22254$'s posterior had less degeneracy around its expectation, it had relatively heavier ``wings'', as compared to $\#24600$. This interpretation is reinforced by the fact that \#24600 started out with a more informed uniform distribution, as compared to \#22254, but still maintained a higher relative entropy despite this. 

\section{Conclusions}

Research in the area of Bayesian equilibrium reconstruction has rapidly advanced since the work of Svesson and Werner\cite{svensson2007,svensson2008}, which has gone from analytic inversion leveraging very few physical assumptions to the current state-of-the-art where complex force-balance models can be seamlessly folded into a non-analytic, robust inference on over 1000 model parameter dimensions. Today, Bayesian equilibrium reconstruction compensates for broken diagnostics \emph{in situ}\cite{vonnessi2012, vonnessi2013} in addition to being able to marginalise out uncertainties due to conducting surface currents, all while preserving the integrity of the inference results. This paper presents the most recent advancements in the area, which surround the computational aspects of analysing the posterior. The end result being that the equilibrium for a high-performance MAST discharge has been shown to be consistent with static GS force-balance, implying that the current selection of diagnostics used in this analysis will need to be expanded, if one wishes to resolve physics not already represented in the GS equation.

Developing research endeavours in this area include adding in a toroidal flow component into the force-balance relation, along with more diagnostic data, and seeing how this affects the inference on MAST discharges. Work is also progressing on deploying BEAST on the KSTAR experiment, where both 2D MSE and diamagnetic loop data can be leveraged to better constrain the equilibrium inference. On the computational end of research, the possibility of deploying machine learning techniques to generate better initial guesses for posterior optimisation is being explored, as it is now the search for the MAP which takes up the majority of computational time (as opposed to the construction of posterior quadratures).

\appendix
\section{The Screening Mitigation Optimiser}\label{SMO3}

In this section we briefly outline the directional search algorithm developed for use in BEAST's optimisation of the posterior. The direction search starts from an initial guess, $x_0$, a given, scalar increment, $\delta$, and proceeds as follows.
\begin{enumerate}
\item At $x_0$ the target function is evaluated, with the value stored (denoted $f_0$).
\item If $\delta$ is smaller than a pre-defined threshold, the algorithm terminates. Otherwise, the procedure continues onto the next step.
\item Evaluate the target function at $x_0\pm \delta e_i$, where $e_i$ is the unit vector for the $i$th coordinate, for all coordinate directions. Sign/direction combinations showing no improvement over $f_0$ (in the case of the posterior, are less than $f_0$) are discarded, with all other combinations being recorded and ranked according to which ones gave the largest improvement over $f_0$. We label each improving coordinate increment as $g_i$, with lower indices having greater improvement over $f_0$; i.e. $g_i$ will generally have the form $\pm\delta e_j$, with $i$ and $j$ uncorrelated. If no direction is found that improves $f_0$, the value of $\delta$ is scaled down (in BEAST $\delta$ is scaled down by a factor of $2$) and the algorithm returns to step 2.
\item Count the number of $g_i$'s and record this value as $M$.
\item Evaluate the target at
\begin{eqnarray*}
x_1:=x_0+\sum_{i=1}^M g_i.
\end{eqnarray*}
\item If the evaluation at $x_1$ produces a result better than $f_0$, a line search is performed along $x_1-x_0$ from the point $x_0$. The result of this search replaces the value of $x_0$ and the algorithm returns to step 1, with $\delta$ being set to it's initial value. Otherwise, $M$ is decremented by one and the algorithm returns to step 5.
\end{enumerate}

In BEAST, a golden section\cite{kiefer1953} line search is used in the above; but any line search method could be applied. The key point to tho above approach in that it is a ``breadth-first'' algorithm in that it will try to change as many model parameter coordinates as possible in each step, as opposed to accepting possibly better gains by moving along just a few coordinates. Indeed, moving along one coordinate at any given step, may indeed produce a better immediate result; but this has a tendency to drive the optimiser into local maxima presented by the screening solutions discussed in \secref{SMO2}. This is the same problem has also been found with both steepest descent and conjugate gradient optimisers, when deployed in BEAST. This last point is unsurprising, as one wouldn't expect such algorithms to be effective on functions with many local maxima. The above algorithm is designed specifically to avoid these local maxima and has proven to be extremely robust in BEAST inferences and has the added advantage that it does not require gradient calculations.

\section{Stochastic Lebesgue Quadrature}\label{SLQ}
One can argue that all of Bayesian inference can be reduced to posterior quadrature calculations. Indeed, any statistical moments of model parameters or marginalisations thereof can be represented as
\begin{eqnarray}
\mathcal{I}:=\int_{\mathbb{R}^n}\mathcal{Q}(\vlambda)\,d\vlambda,\label{slq1}
\end{eqnarray}
where $n$ being the number of model parameters, with diagnostic observations implicitly held as parameters within $\mathcal{Q}(\vlambda)$. With this, we seek to develop a numerical scheme to integrate $\mathcal{I}$ for possibly large values of $n$.

First, we assume $\mathcal{I}<\infty$, to guarantee the existence of a bounded set $\mathcal{B}$ such that
\begin{eqnarray}
\left|\mathcal{I}-\int_{\mathcal{B}}\mathcal{Q}(\vlambda)\,d\vlambda\right|<\epsilon,\label{slq2}
\end{eqnarray}
for any given $\epsilon > 0$. Thus, we are able to make the following approximation
\begin{eqnarray}
\mathcal{I} & \approx & \int_{\mathcal{B}}\mathcal{Q}(\vlambda)\,d\vlambda\nonumber\\
& = & \int_\mathcal{B}\int_0^{\mathcal{Q}(\vlambda)}\ dtd\vlambda\nonumber\\
& = & \int_{\mathbb{R}^n}\int_0^{\mathcal{Q}(\vlambda)}|\mathcal{B}|\mathcal{U}_\mathcal{B}(\vlambda)\,dtd\vlambda\nonumber\\
& = & |\mathcal{B}|\int_0^\infty\int_{\left\{\vlambda\,|\,\mathcal{Q}(\vlambda)>t\right\}}\mathcal{U}_\mathcal{B}(\vlambda)\,d\vlambda dt,\label{slq3}
\end{eqnarray}
where $|\mathcal{B}|$ denotes the $n$-dimensional volume of $\mathcal{B}$ and
\begin{eqnarray*}
\mathcal{U}_\mathcal{B}(\vlambda):=\left\{\begin{array}{rl}|\mathcal{B}|^{-1}, & \vlambda\in\mathcal{B}\\0, & \text{otherwise}\end{array}\right..
\end{eqnarray*}
One will note that the co-area formula\cite{federer2006} has been employed in the last step of \eqref{slq3}. It is clear that $\mathcal{U}_\mathcal{B}(\vlambda)$ is normalised to one by definition and thus, constitutes a uniform probability distribution. The motivation for this particular factorisation is embodied in the following definition
\begin{eqnarray}
\xi(t) & := &  \int_{\left\{\vlambda\,|\,\mathcal{Q}(\vlambda)>t\right\}}\mathcal{U}_\mathcal{B}(\vlambda)\,d\vlambda\nonumber\\
& = & \text{Proportion of $\mathcal{B}$ where $\mathcal{Q}(\vlambda)>t$},\label{slq4}
\end{eqnarray}
which hints at a way in which uniform sampling may be employed to obtain the desired integral. To fully realise this, we note \eqref{slq4} directly indicates that $\xi(t)$ may be statistically inverted via ordering uniform samples of $\mathcal{B}$ with respect to their $\mathcal{Q}$ evaluations. Indeed, given a collection of $m$ uniform samples of $\mathcal{B}$, denoted $\vlambda_i$, indexed according to 
\begin{eqnarray}
\mathcal{Q}(\vlambda_1)>\mathcal{Q}(\vlambda_2)>\cdots>\mathcal{Q}(\vlambda_m),\label{slq5}
\end{eqnarray}
then \eqref{slq4} indicates that
\begin{eqnarray}
\xi^{-1}\left(\frac{i}{m}\right) \approx \mathcal{Q}_i := \mathcal{Q}(\vlambda_i).\label{slq6}
\end{eqnarray}
To use this insight, we employ the definition in \eqref{slq4} and make the substitution $v=\xi(t)$ to reduce the expression in \eqref{slq3}:
\begin{eqnarray}
\mathcal{I} & \approx & \int_0^\infty|\mathcal{B}|\xi(t)\,dt\nonumber\\
& = & |\mathcal{B}|\int_1^0v\xi^{-1\prime}(v)\,dv\nonumber\\
& = & |\mathcal{B}|\left[v\xi^{-1}(v)\Big|_1^0+\int_0^1\xi^{-1}(v)\,dv\right]\nonumber\\
& = & |\mathcal{B}|\left[-\inf_\mathcal{\vlambda\in B}\mathcal{Q(\vlambda)}+\int_0^1\xi^{-1}(v)\,dv\right]\nonumber\\
& \approx & |\mathcal{B}|\int_0^1\xi^{-1}(v)\,dv,\label{slq7}
\end{eqnarray}
where we have used integration by parts and the assumption that $\mathcal{B}$ can be chosen to make $\inf_\mathcal{\vlambda\in B}\mathcal{Q(\vlambda)}$ small enough to satisfy the desired level of accuracy for the quadrature. One will note that the assumption of $\mathcal{I}<\infty$ automatically implies that $\lim_{v\to 0}v\xi^{-1}(v)=0$. 

While the above is very similar to the development presented in von Nessi \etal\cite{vonnessi2013}, it differs from that derivation in that the quadrature transformation has no intrinsic reliance on the demarkations of likelihoods and priors. Indeed, the above result is quite general in that $\mathcal{Q}(\vlambda)$ need not be related to a probability distribution. Moreover, the definition of $\xi(t)$ needed to be altered to accommodate the transform's reliance on uniform distributions, which ultimately leads to the addition of the $|\mathcal{B}|$ term in the final expression.

\subsection*{Generalisations of Nested Sampling}
The expression in \eqref{slq6} indicates that a graph of $\xi(t)$ can be statistically constructed by taking an ordered set of $\mathcal{Q}$ evaluations coming from $m$ uniform samples of $\mathcal{B}$ with an abscissa constructed of $m$ ordered uniform samples taken on $[0,1]$. Concretely, given the $m$ samples denoted in \eqref{slq6}, one can take $m$ uniform samples, $t_i$, from $[0,1]$ indexed so that $1\ge t_m> t_{m-1}>\cdots>t_1\ge0$ to approximate
\begin{eqnarray}
\xi^{-1}\left(t_i\right) \approx \mathcal{Q}_i.\label{slq8}
\end{eqnarray}
One can refine the resolution of any part of this graph by adding, say $k$, abscissa values coming from uniform samples on $[t_i, t_j]$ and $k$ values coming from uniform samples taken from $\left\{\vlambda\in\mathcal{B}\,|\,\mathcal{Q}_j<\mathcal{Q}(\vlambda)<\mathcal{Q}_i\right\}$ and reordering both sets of $k+m$ values according to \eqref{slq5}--\eqref{slq8}. This is a general prescription for graph refinement, for which Skilling's NS is a particular instance of.\cite{sivia2006,skilling2006,vonnessi2013} Indeed, this method can be directly leveraged to design both single-threaded and multi-threaded generalisations of NS. However, for the results presented in this paper, we retain the original NS methodology for refining the graph in \eqref{slq8}, which is detailed elsewhere\cite{sivia2006,skilling2006,vonnessi2013}. Once a refinement of sufficient accuracy has been achieved, $\mathcal{I}$ is simply evaluated via \eqref{slq7}, \eqref{slq8} and the application of a trapezoidal quadrature rule.

\subsection*{Uniform Sampling}
The computational tractability of SLQ relies on the ability to generate uniform samples from the set $\mathcal{B}_t:=\left\{\vlambda\in\mathcal{B}\,|\,t<\mathcal{Q}(\vlambda)\right\}$ for any $0\le t\le t^*:=\sup_{\vlambda\in\mathcal{B}}\mathcal{Q}(\vlambda)$ efficiently. Clearly, \emph{ab initio} uniform sampling of $\mathcal{B}$ will be rendered unacceptably inefficient as a proxy for sampling on $\mathcal{B}_t$ for values of $t$ approaching $t^*$. One approach to dealing with this is to approximate $\mathcal{B}_t$ by a collection of $n$ dimensional, pair-wise disjoint hypercubes, denoted $\mathcal{C}_i$, each with their coordinate axis corresponding to the collection of model parameters in the problem. These cubes need not be of the same volume or proportion. Each cube is then assigned a relative probability based on its volume:
\begin{eqnarray}
\mathcal{P}(\mathcal{C}_i):= \frac{|\mathcal{C}_i|}{\left|\bigcup_i\mathcal{C}_i\right|}.\label{slq9}
\end{eqnarray}
To uniformly sample from $\bigcup_i\mathcal{C}_i$, one first needs to randomly select a particular hypercube's index according to the probability in \eqref{slq9}, then one can perform uniform Gibb's sampling on the selected cube to finally generate the next uniform sample from $\bigcup_i\mathcal{C}_i\approx \mathcal{B}_t$. Indeed, this prescription can be viewed as a Gibb's sampling over $n$ dimensions, plus one discretised dimension corresponding to the indexing on the hypercube approximation. This sampling over a union of hypercubes can be carried out very quickly, even in high-dimensions with many cubes in the collection; and thus offers an appealing foundation on which to build a statistical quadrature.

The next point to be addressed is how to create and maintain a collection of hypercubes which closely approximates $\mathcal{B}_t$. To this end a collection of points $s_i\in \mathcal{B}_t$ is first created and evolved to directly correspond to the pool of samples used in the NS quadrature construction. In addition to these points, the MAP and a collection of \emph{ab initio} uniform samples from $\mathcal{B}$ are initially added to the collection of $s_i$. Once any of these points fail to meet the $\mathcal{Q}$ constraint in the NS progression, Skilling's multi-state leapfrog algorithm (see \S30.4 in MacKay\cite{mackay2003} for details) is employed on the collection of $s_i$ to find new points to replace those that no longer meet the $\mathcal{Q}$-constraint. If a fixed number of attempts fails to produce points that satisfy the new $\mathcal{Q}$-constraint, then effort is abandoned and the collection of $s_i$ is reduced accordingly.

Once the collection of $s_i$ has been established, the construction of the hypercubes proceeds as follows:
\begin{enumerate}
\item Establish a minimal hypercube with axes corresponding to model parameters in the inference which contains all $s_i$. We denote this hypercube $\mathcal{C}_g$.
\item For each $s_i$ create a hypercube, $\mathcal{C}_i$ of the same size and orientation as $\mathcal{C}_g$.
\item Perform a pairwise comparison between all $\mathcal{C}_i$, going through each dimensions to see if they are disjoint. Note that the cubes need only be non-overlapping in one dimension to be disjoint. 
\begin{enumerate}
\item If a pair of hypercubes, $\mathcal{C}_i$ and $\mathcal{C}_j$, are overlapping, establish the coordinate that constitutes their greatest separation. We will label this the $k$th coordinate for convenience.
\item If both $\mathcal{C}_i$ and $\mathcal{C}_j$ have their current bounding hyperplane along the $k$th coordinate lying between $s_i$ and $s_j$, these hyperplanes are adjusted to have their $k$th coordinate be the average of the $k$th coordinates of both their previous positions.
\item If only one of $\mathcal{C}_i$ or $\mathcal{C}_j$ has their current bounding hyperplane along the $k$th coordinate lying between $s_i$ and $s_j$, then the other's bounding hyperplane's $k$th coordinate is adjusted to coincide with that of the one that separates $s_i$ and $s_j$.
\item If neither  $\mathcal{C}_i$ nor $\mathcal{C}_j$ has their current bounding hyperplane along the $k$th coordinate lying between $s_i$ and $s_j$, then then both bounding hyperplanes are adjusted so their $k$th coordinate coincides with the $k$th coordinate of the average position between $s_i$ and $s_j$.
\end{enumerate}
\end{enumerate}
Once the collection of approximating hypercubes is first established, changes, additions and removals (corresponding to the evolution of $s_i$) can be made in accordance with the above pseudo-code in $O(m)$ time, where $m$ is the current number of cubes. This is achieved primarily by tracking the coordinates along which each pair of cubes is separated and by noting which bounding hyperplanes correspond to those of $\mathcal{C}_g$ for each $\mathcal{C}_i$ and subsequently using this information to minimise the number of comparisons made on hypercube insertions and deletions.

The above algorithm ensures that $\bigcup_i\mathcal{C}_i$ contains all $s_i\in\mathcal{C}_g$ with no overlaps between cubes; although this union will not be a cover for $\mathcal{C}_g$, in general. As $\mathcal{C}_g$ will typically be a poor approximation of $\mathcal{B}_t$, a global scaling factor, $\eta$, along with a family of linear mappings $L_\eta$ on the collection of $\mathcal{C}_i$, having the properties $L_0(\mathcal{C}_i) = \{s_i\}$ and $L_1(\mathcal{C}_i) = \mathcal{C}_i$ is also introduced. Ultimately, new uniform samples are drawn from
\begin{eqnarray}
\bigcup_iL_\eta(\mathcal{C}_i)\approx \mathcal{B}_t,\label{slq10}
\end{eqnarray}
where $\eta$ is dynamically adjusted on the interval $[1,0)$ to achieve a desired level of efficiency for uniform samples having $\mathcal{Q}$-evaluations greater than $t$.

\begin{acknowledgments}
This work was jointly funded by the Australian Government through International Science Linkages Grant CG130047, the Australian Research Council Grant FT0991899, the Australian National University, the United Kingdom Engineering and Physical Sciences Research Council under grant EP/G003955, and by the European Communities under the contract of Association between EURATOM and CCFE. The views and opinions expressed herein do not necessarily reflect those of the European Commission.
\end{acknowledgments}


\begin{thebibliography}{10}

\bibitem{goedbloed2004}
J.~P. Goedbloed and S.~Poedts.
\newblock {\em {Principles of Magnetohydrodynamics: With Applications to
  Laboratory and Astrophysical Plasmas}}.
\newblock Cambridge University Press, August 2004.

\bibitem{hutchinson2005}
I.~H. Hutchinson.
\newblock {\em {Principles of Plasma Diagnostics}}.
\newblock Cambridge University Press, second edition, July 2005.

\bibitem{bellan2008}
P.~M. Bellan.
\newblock {\em {Fundamentals of Plasma Physics}}.
\newblock Cambridge University Press, first edition, September 2008.

\bibitem{goedbloed2010}
J.~P. Goedbloed, R.~Keppens, and S.~Poedts.
\newblock {\em {Advanced Magnetohydrodynamics: With Applications to Laboratory
  and Astrophysical Plasmas}}.
\newblock Cambridge University Press, first edition, June 2010.

\bibitem{wesson2011}
J.~Wesson.
\newblock {\em {Tokamaks}}.
\newblock International Series of Monographs on Physics. Oxford University
  Press, USA, fourth edition, December 2011.

\bibitem{ferron1998}
J.~R. Ferron, M.~L. Walker, L.~L. Lag, H.~E. St.~John, D.~A. Humphreys, and
  J.~A. Leuer.
\newblock {Real Time Equilibrium Reconstruction for Tokamak Discharge Control}.
\newblock {\em Nuclear Fusion}, 38(7):1055--1066, 1998.

\bibitem{mcardle2004}
G.~J. Mcardle and J.~Storrs.
\newblock {First results from the MAST digital plasma control system}.
\newblock {\em Fusion Engineering and Design}, 71(1-4):59--64, June 2004.

\bibitem{mazon2010}
D.~Mazon, J.~Blum, C.~Boulbe, B.~Faugeras, A.~Boboc, M.~Brix, P.~De-Vries,
  S.~E. Sharapov, and L.~Zabeo.
\newblock {Equinox: A Real-Time Equilibrium Code and Its Validation at JET}.
\newblock In {\em 4th International Conference on Physics and Control},
  Catania, Italy, 2010. World Scientific.

\bibitem{blum2012}
J.~Blum, C.~Boulbe, and B.~Faugeras.
\newblock {Reconstruction of the Equilibrium of the Plasma in a Tokamak and
  Identification of the Current Density Profile in Real Time}.
\newblock {\em Journal of Computational Physics}, 231(3):960--980, February
  2012.

\bibitem{kurihara2000}
K.~Kurihara.
\newblock {A New Shape Reproduction Method Based on the Cauchy-Condition
  Surface for Real-Time Tokamak Reactor Control}.
\newblock {\em Fusion Engineering and Design}, 51:1049--1057, 2000.

\bibitem{kurihara2005}
K.~Kurihara, Y.~Kawamata, M.~Sueoka, H.~Hosoyama, I.~Yonekawa, T.~Suzuki,
  T.~Oikawa, and S.~Ide.
\newblock {The Basic Methods for Understanding of Plasma Equilibrium Toward
  Advanced Control}.
\newblock {\em Fusion Engineering and Design}, 74(1-4):527--536, November 2005.

\bibitem{lao1985}
L.~L. Lao, H.~E. St~John, R.~D. Stambaugh, A.~G. Kellmann, and W.~Pfeiffer.
\newblock {Reconstruction of Current Profile Parameters and Plasma Shapes in
  Tokamaks}.
\newblock {\em Nuclear Fusion}, 25(11):1611--1622, 1985.

\bibitem{lao1985b}
L.~L. Lao, H.~E. St.~John, R.~D. Stambaugh, and W.~Pfeiffer.
\newblock {Separation of Beta-Barp and Li in Tokamaks of Non-Circular
  Cross-Section}.
\newblock {\em Nuclear Fusion}, 25(10):1421--1436, 1985.

\bibitem{zwingmann2003}
W.~Zwingmann.
\newblock {Equilibrium analysis of steady state tokamak discharges}.
\newblock In {\em Nuclear Fusion}, pages 842--850. CEA Cadarache, EURATOM
  Assoc, DSM, DRFC, F-13108 St Paul Les Durance, France, 2003.

\bibitem{fitzgerald2013}
M.~Fitzgerald, L.~C. Appel, and M.~J. Hole.
\newblock {EFIT Tokamak Equilibria with Toroidal Flow and Anisotropic Pressure
  Using the Two-Temperature Guiding-Centre Plasma}.
\newblock {\em Nuclear Fusion}, 53(11):113040, October 2013.

\bibitem{cottrell1991}
G.~A. Cottrell.
\newblock {\em {Maximum Entropy in Action}}, chapter {Maximum Entropy and
  Plasma Physics}.
\newblock Oxford University Press, September 1991.

\bibitem{millar2000}
A.~P. Millar, D.~C. McDonald, and D.~A. Diver.
\newblock {Genetic Algorithms in Plasma Diagnostic Analysis}.
\newblock {\em Plasma Physics and Controlled Fusion}, 42(3):337--346, 2000.

\bibitem{fischer2003}
R.~Fischer, A.~Dinklage, and E.~Pasch.
\newblock {Bayesian Modelling of Fusion Diagnostics}.
\newblock {\em Plasma Physics and Controlled Fusion}, 45(7):1095--1111, 2003.

\bibitem{dinklage2004}
A.~Dinklage, R.~Fischer, and J.~Svensson.
\newblock {Topics and Methods for Data Validation by Means of Bayesian
  Probability Theory}.
\newblock {\em Fusion Science and Technology}, 46(2):355--364, 2004.

\bibitem{svensson2004}
J.~Svensson, A.~Dinklage, J.~Geiger, A.~Werner, and R.~Fischer.
\newblock {Integrating Diagnostic Data Analysis for W7-AS Using Bayesian
  Graphical Models}.
\newblock {\em Review of Scientific Instruments}, 75(10):4219, 2004.

\bibitem{svensson2008}
J.~Svensson, A.~Werner, and {JET-EFDA Contributors}.
\newblock {Current Tomography for Axisymmetric Plasmas}.
\newblock {\em Plasma Physics and Controlled Fusion}, 50(8):085002, May 2008.

\bibitem{ford2008}
O.~Ford, J.~Svensson, A.~Boboc, D.~C. McDonald, and {the JET-EFDA
  Contributors}.
\newblock {Forward Modeling of JET Polarimetry Diagnostic}.
\newblock {\em Review of Scientific Instruments}, 79(10):10F324, 2008.

\bibitem{reginatto2008}
M.~Reginatto and A.~Zimbal.
\newblock {Bayesian and Maximum Entropy Methods for Fusion Diagnostic
  Measurements with Compact Neutron Spectrometers}.
\newblock {\em Review of Scientific Instruments}, 79(2):023505, 2008.

\bibitem{dose2003}
V.~Dose.
\newblock {Bayesian Inference in Physics: Case Studies}.
\newblock {\em Reports on Progress in Physics}, 66(9):1421, 2003.

\bibitem{vontoussaint2011}
U.~von Toussaint.
\newblock {Bayesian Inference in Physics}.
\newblock {\em Reviews of Modern Physics}, 83(3):943--999, September 2011.

\bibitem{enting1995}
I.~G. Enting, C.~M. Trudinger, and R.~J. Francey.
\newblock {A Synthesis Inversion of the Concentration and $\delta^13$C of
  Atmospheric CO$_2$}.
\newblock {\em Tellus B}, 47B:35--52, 1995.

\bibitem{wang2009}
Y.-P. Wang, C.~M. Trudinger, and I.~G. Enting.
\newblock {A Review of Applications of Model--Data Fusion to Studies of
  Terrestrial Carbon Fluxes at Different Scales}.
\newblock {\em Agricultural and Forest Meteorology}, 149(11):1829--1842,
  November 2009.

\bibitem{wang2011}
Y.-P. Wang, E.~Kowalczyk, R.~Leuning, G.~Abramowitz, M.~R. Raupach, B.~Pak,
  E.~van Gorsel, and A.~Luhar.
\newblock {Diagnosing Errors in a Land Surface Model (CABLE) in the Time and
  Frequency Domains}.
\newblock {\em Journal of Geophysical Research}, 116(G1):G01034, March 2011.

\bibitem{haverd2013}
V.~Haverd, M.~R. Raupach, P.~R. Briggs, J.~G. Canadell, P.~Isaac,
  C.~Pickett-Heaps, S.~H. Roxburgh, E.~van Gorsel, R.~A. Viscarra~Rossel, and
  Z.~Wang.
\newblock {Multiple Observation Types Reduce Uncertainty in Australia's
  Terrestrial Carbon and Water Cycles}.
\newblock {\em Biogeosciences}, 10(3):2011--2040, 2013.

\bibitem{raupach2005}
M.~R. Raupach, P.~J. Rayner, D.~J. Barrett, R.~S. DeFries, M.~Heimann, D.~S.
  Ojima, S.~Quegan, and C.~C. Schmullius.
\newblock {Model-Data Synthesis in Terrestrial Carbon Observation: Methods,
  Data Requirements and Data Uncertainty Specifications}.
\newblock {\em Global Change Biology}, 11(3):378--397, March 2005.

\bibitem{hole2010}
M.~J. Hole, G.~T. von Nessi, J.~Bertram, J.~Svensson, L.~C. Appel, B.~D.
  Blackwell, R.~L. Dewar, and J.~Howard.
\newblock {Model Data Fusion: Developing Bayesian Inversion to Constrain
  Equilibrium and Mode Structure}.
\newblock {\em Journal of Plasma and Fusion Research}, 9:479--486, 2010.

\bibitem{hole2010a}
M.~J. Hole, G.~T. von Nessi, D.~G. Pretty, J.~Howard, B.~D. Blackwell,
  J.~Svensson, and L.~C. Appel.
\newblock {The Use of Bayesian Inversion to Resolve Plasma Equilibrium}.
\newblock {\em Review of Scientific Instruments}, 81(10):10E127, 2010.

\bibitem{hole2011a}
M.~J. Hole, G.~T. von Nessi, J.~Svensson, L.~C. Appel, and {the MAST team}.
\newblock {An Equilibrium Validation Technique Based on Bayesian Inference}.
\newblock {\em Nuclear Fusion}, 51(10):103005, August 2011.

\bibitem{vonnessi2012}
G.~T. von Nessi, M.~J. Hole, J.~Svensson, and L.~C. Appel.
\newblock {Evidence Cross-Validation and Bayesian Inference of MAST Plasma
  Equilibria}.
\newblock {\em Physics of Plasmas}, 19(1):012506, 2012.

\bibitem{vonnessi2013}
G.~T. von Nessi, M.~J. Hole, and {the MAST team}.
\newblock {A Unified Method for Inference of Tokamak Equilibria and Validation
  of Force-Balance Models Based on Bayesian Analysis}.
\newblock {\em Journal of Physics A: Mathematical and Theoretical},
  46(18):185501, April 2013.

\bibitem{ford2010}
O.~Ford.
\newblock {\em {Tokamak Plasma Analysis through Bayesian Diagnostic
  Modelling}}.
\newblock PhD thesis, University of London, November 2010.

\bibitem{jaynes2003}
E.~T. Jaynes.
\newblock {\em {Probability Theory: The Logic of Science}}.
\newblock Cambridge University Press, June 2003.

\bibitem{sivia2006}
D.~Sivia and J.~Skilling.
\newblock {\em {Data Analysis: A Bayesian Tutorial}}.
\newblock Oxford University Press, USA, second edition, July 2006.

\bibitem{svensson2007}
J.~Svensson and A.~Werner.
\newblock {Large Scale Bayesian Data Analysis for Nuclear Fusion Experiments}.
\newblock In {\em 2007 IEEE International Symposium on Intelligent Signal
  Processing (WISP 2007)}, pages 1--6, October 2007.

\bibitem{mackay2003}
D.~J.~C. MacKay.
\newblock {\em {Information Theory, Inference and Learning Algorithms}}.
\newblock Cambridge University Press, first edition, October 2003.

\bibitem{hooke1961}
R.~Hooke and T.~A. Jeeves.
\newblock {``Direct Search" Solution of Numerical and Statistical Problems}.
\newblock {\em Journal of the ACM}, 8(2):212--229, 1961.

\bibitem{skilling2006}
J.~Skilling.
\newblock {Nested Sampling for General Bayesian Computation}.
\newblock {\em Bayesian Analysis}, 1(4):833--859, 2006.

\bibitem{gilbarg2001}
D.~Gilbarg and N.~S. Trudinger.
\newblock {\em {Elliptic Partial Differential Equations of Second Order}}.
\newblock Springer, 1998 edition, January 2001.

\bibitem{kiefer1953}
J.~Kiefer.
\newblock {Sequential Minimax Search for a Maximum}.
\newblock {\em Proceedings of the American Mathematical Society},
  4(3):502--506, 1953.

\bibitem{federer2006}
H.~Federer.
\newblock {\em {Geometric Measure Theory}}.
\newblock {Springer-Verlag}, 2006.

\end{thebibliography}
\end{document}